\documentstyle[psfig,epsf]{mn}

\newif\ifAMStwofonts

\ifoldfss
  \ifCUPmtlplainloaded \else
    \NewTextAlphabet{textbfit} {cmbxti10} {}
    \NewTextAlphabet{textbfss} {cmssbx10} {}
    \NewMathAlphabet{mathbfit} {cmbxti10} {} 
    \NewMathAlphabet{mathbfss} {cmssbx10} {} 
  \fi
  \ifAMStwofonts
    \ifCUPmtlplainloaded \else
      \NewSymbolFont{upmath} {eurm10}
      \NewSymbolFont{AMSa} {msam10}
      \NewMathSymbol{\upi}     {0}{upmath}{19}
      \NewMathSymbol{\umu}     {0}{upmath}{16}
      \NewMathSymbol{\upartial}{0}{upmath}{40}
      \NewMathSymbol{\leqslant}{3}{AMSa}{36}
      \NewMathSymbol{\geqslant}{3}{AMSa}{3E}

    \fi
  \fi
\fi 

\ifnfssone
  \newmathalphabet{\mathit}
  \addtoversion{normal}{\mathit}{cmr}{m}{it}
  \addtoversion{bold}{\mathit}{cmr}{bx}{it}
  \newmathalphabet{\mathbfit} 
  \addtoversion{normal}{\mathbfit}{cmr}{bx}{it}
  \addtoversion{bold}{\mathbfit}{cmr}{bx}{it}
  \newmathalphabet{\mathbfss} 
  \addtoversion{normal}{\mathbfss}{cmss}{bx}{n}
  \addtoversion{bold}{\mathbfss}{cmss}{bx}{n}
  \ifAMStwofonts
    \ifCUPmtlplainloaded \else
      \UseAMStwoboldmath
      \makeatletter
      \new@mathgroup\upmath@group
      \define@mathgroup\mv@normal\upmath@group{eur}{m}{n}
      \define@mathgroup\mv@bold\upmath@group{eur}{b}{n}
      \edef\UPM{\hexnumber\upmath@group}
      \new@mathgroup\amsa@group
      \define@mathgroup\mv@normal\amsa@group{msa}{m}{n}
      \define@mathgroup\mv@bold\amsa@group{msa}{m}{n}
      \edef\AMSa{\hexnumber\amsa@group}
      \makeatother
      \mathchardef\upi="0\UPM19
      \mathchardef\umu="0\UPM16
      \mathchardef\upartial="0\UPM40
      \mathchardef\leqslant="3\AMSa36
      \mathchardef\geqslant="3\AMSa3E
    \fi
  \fi
\fi 

\ifnfsstwo
  \DeclareMathAlphabet{\mathbfit}{OT1}{cmr}{bx}{it}
  \SetMathAlphabet\mathbfit{bold}{OT1}{cmr}{bx}{it}
  \DeclareMathAlphabet{\mathbfss}{OT1}{cmss}{bx}{n}
  \SetMathAlphabet\mathbfss{bold}{OT1}{cmss}{bx}{n}
  \ifAMStwofonts
    \ifCUPmtlplainloaded \else
      \DeclareSymbolFont{UPM}{U}{eur}{m}{n}
      \SetSymbolFont{UPM}{bold}{U}{eur}{b}{n}
      \DeclareSymbolFont{AMSa}{U}{msa}{m}{n}
      \DeclareMathSymbol{\upi}{0}{UPM}{"19}
      \DeclareMathSymbol{\umu}{0}{UPM}{"16}
      \DeclareMathSymbol{\upartial}{0}{UPM}{"40}
      \DeclareMathSymbol{\leqslant}{3}{AMSa}{"36}
      \DeclareMathSymbol{\geqslant}{3}{AMSa}{"3E}
    \fi
  \fi
\fi 

\ifCUPmtlplainloaded \else
  \ifAMStwofonts \else 
    \def\upi{\pi}
    \def\umu{\mu}
    \def\upartial{\partial}
  \fi
\fi

\title[Near-infrared imaging polarimetry of dusty young stars] {Near-infrared
  imaging polarimetry of dusty young stars} \author[Hales et al.] {A. S. Hales$^1$, T. M. Gledhill$^2$, M. J. Barlow$^1$
  and K.T.E. Lowe$^2$ \\ $^1$Department of Physics and Astronomy,
  University College London, Gower Street, London WC1E 6BT\\
  $^2$School of Physics, Astronomy and Mathematics, University of Hertfordshire, College Lane, Hatfield, AL10 9AB
  }
  \pagerange{\pageref{firstpage}--\pageref{lastpage}} \pubyear{2004}

\def\LaTeX{L\kern-.36em\raise.3ex\hbox{a}\kern-.15em
    T\kern-.1667em\lower.7ex\hbox{E}\kern-.125emX}

\begin{document}

\label{firstpage}

\maketitle

\begin{abstract}

We have carried out JHK polarimetric observations of eleven dusty
young stars, by using the polarimeter module IRPOL2 with the near-IR
camera UIST on the 3.8-m United Kingdom Infrared Telescope
(UKIRT). Our sample targeted systems for which UKIRT-resolvable discs
had been predicted by model fits to their spectral energy
distributions. Our observations have confirmed the presence of
extended polarized emission around TW Hya and around
HD~169142. HD~150193 and HD~142666 show the largest polarization
values among our sample, but no extended structure was resolved. By
combining our observations with HST coronographic data from the
literature, we derive the J- and H-band intrinsic polarization radial
dependences of TW Hya's disc. We find the disc's polarizing efficiency
is higher at H than at J, and we confirm that the J- and H-band
percentage polarizations are reasonably constant with radius in the
region between $0.9$ and $1.3$~arcseconds from the star. We find that
the objects for which we have detected extended polarizations are
those for which previous modelling has suggested the presence of
flared discs, which are predicted to be brighter than flat discs and
thus would be easier to detect polarimetrically.


\end{abstract}


\section{Introduction}

The discovery of the Vega-phenomenon in 1984 is considered one of the
most important achievements of the IRAS space telescope. It discovered
that several main sequence (MS) stars, including Vega, exhibited large
mid- and far-infrared excesses that could not be ascribed to pure
photospheric emission \cite{aum84}. Such excesses were attributed to a
disc or ring of solid particles surrounding the stars, later termed
`debris-discs', and great interest was taken in these objects because
of their relevance to the formation of planetary systems.


The case of young low- and intermediate-mass systems, such as T Tauri
and Herbig Ae/Be (HAeBe) stars, is of particular interest
\cite{yudin00,waters98}.
These stars are considered to be precursors of classic Vega-like
stars, such as $\beta$ Pictoris, $\epsilon$ Eridani, HR 4796A,
Fomalhaut and Vega itself. While planetary bodies are thought to have
already formed in Vega-like systems \cite{aum84,beust96,natta04}, the
circumstellar environments of T Tauri and HAeBe stars probably
represent an early phase of planet formation \cite{bouwman01}.

From high resolution optical spectra of a sample of young dusty stars,
the majority in common with our current sample and classified as HAeBe
or T Tauri stars on the basis of their emission line spectra, Dunkin,
Barlow \& Ryan \shortcite{dun97a,dun97b} found that analysis of their
absorption line spectra yielded surface gravities typical of MS stars,
suggesting that they had already arrived on the MS. These type of
objects have been referred to as `transition' objects \cite{malfait98}
or `dusty' Vega-excess stars \cite{dent05} and they are nowadays
thought to play a pivotal role in our understanding on the transition
between the pre-MS (T Tauri and HAeBe stars) and the MS (Vega-like
stars).

Many issues remain unclear regarding the circumstellar structures of these
`transition' objects, in particular, whether they are
surrounded by flat discs or spherical envelopes is still not
determined \cite{vink02}. Based on their spectral energy
distributions, Meeus et al. \shortcite{meeus01} introduced a
classification of HAeBe stars into two groups: Group I objects have an
almost flat or even rising 20-100~$\mu$m spectral energy distribution
(SED), which can be well modelled by a flaring disc \cite{dudu04}. On
the other hand, Group II objects have a much bluer IR SED, and are
believed to be associated with self-shadowed non-flared discs, which
are predicted to be smaller in spatial extent than those of Group I
sources.

High-contrast, high-spatial resolution, imaging close to bright
central stars is required to test the validity of these
models. Despite the efforts of many groups, limitations on
instrumental capabilities have to date kept the number of directly
imaged discs to a very low-number(see Zuckerman \shortcite{zuck01}
for a review).
 When only SEDs are available, not all parameters can be uniquely
determined, e.g. the grain-size distribution and the spatial density
profile are usually not separable.

%

Imaging polarimetry provides a powerful technique for detecting
dust-discs around bright stars. Since only the light from the discs is
expected to be polarized, the bright central stars are automatically
suppressed in polarized light images. The utility of imaging
polarimetry was demonstrated over ten years ago when Gledhill,
Scarrott \& Wolstencroft \shortcite{gled91} detected polarized
emission from the nearby Beta Pictoris disc ($>1000$~AU in radius at a
distance of $19.3$~pc; Crifo et al \shortcite{crifo97}). The light was
found to be 17\% polarized and extended up to 30 arcseconds from the
star (R- waveband). Kuhn, Potter \& Parise \shortcite{kuhn01} also
used imaging polarimetry to detect polarized light around the A5Ve
star HD~169142. More recently, using the Very Large Telescope (VLT),
Apai et al. \shortcite{apai04} tested the power of imaging polarimetry
on an 8-m class telescope, detecting the presence of polarized
emission as close as $0.1$ arcsecond ($\sim6$~AU) from the classical T
Tauri star TW Hya.

We have carried out a JHK imaging polarimetry survey for a sample of
ten `transition' objects (so called `dusty' Vega-excess stars), selected from
the surveys of Sylvester et al. \shortcite{syl96} and Mannings \&
Barlow \shortcite{mann98}, along with TW Hya. Modelling of their SEDs
had predicted angular sizes that could be resolvable$\;$ at
near-infrared (near-IR) wavelengths using the near-IR imaging
instrument UIST on the 3.8-m United Kingdom Infrared Telescope
\cite{sylskin96,syl97}. Our primary aim is to detect the signature of
extended polarized emission from a scattering disc around the target
stars. Note that even if the disc is too small to be resolved with
UKIRT, a net polarization can be observed, indicating the presence and
orientation of a disc on the sky.

Section 1.1 describes our target sample and basic selection
criteria. $\S2$ reports on the observations and $\S3$ on the data
reduction; Results are presented in $\S4$, and discussed in $\S5$.

\subsection{Target sample}

We have selected a sample of ten young stars with dust discs from the
surveys of Sylvester et al. \shortcite{syl96} and Mannings \& Barlow
\shortcite{mann98}. All targets show mid- and far-IR excesses due to
warm and cold dust grains. Most of them have been thoroughly studied
using spectrophotometric techniques. Spectral types, effective
temperatures, surface gravities, elemental abundances and overall
line-emission characteristics are available for the majority of the
sample \cite{dun97a,dun97b}. A summary of our sample is shown in
Table~\ref{sample}, where we have listed respectively their names,
spectral classification, infrared excess fractions L$_{\rm
IR}$/L$_{\rm \star}$, distances and V-band magnitudes (as listed in
the SIMBAD database).


The IR excesses of our targets are all much larger than those of true
Vega-like stars ( which are typically L$_{\rm IR}$/L$_{\rm
\star}<10^{-3}$ for Vega-like stars; Sylvester et
al. \shortcite{syl96}). Their surface gravities indicate they have
recently reached the MS \cite{dun97a,dun97b}. HD~123160, HD~123356,
HD~141569 and HD~145263 have relatively low IR excesses compared to
the rest of our sample. The remaining stars have considerably larger
IR excesses (L$_{\rm IR}$/L$_{\rm \star}>0.1$), suggesting that they
might be in a somewhat younger evolutionary state. HD~142666,
HD~143006, HD~144432 and HD~169142 have near-IR excesses attributable
to hot dust emission, resulting in (J-H) and (H-K) colours coincident
with the locus of HAeBe and T Tauri stars \cite{syl96}. It has been
proposed that the dust responsible for the near-IR excesses is a main
agent for the optical polarization, and thus that Vega-like
`prototypes' showing almost no near-IR excess should present low
polarization values \cite{yudin00}. HD~142666, HD~143006, HD~144432
and HD~169142 have been detected in CO in the survey by Dent et
al. \shortcite{dent05}, indicating they still have substantial gas
content.

\section{Observations}

The J-band data presented in this work was acquired on April 29th
2003, using the dual-beam polarimeter module IRPOL2 in conjunction
with the near-IR camera UIST on UKIRT at Mauna Kea observatory. The
resulting field of view was of 15$\times$60~arcseconds, with 0.12
arcseconds per pixel. H- and K-band data for TW Hya, HD~150193 and
HD~141569 were also acquired, on April 30th 2003, using the same
observing procedure.

Dual-beam imaging polarimetry was obtained using a lithium niobate
Wollaston prism, along with a rotating half-wave plate. By observing
at four different half-wave plate positions (separated by
22.5~degrees), linear polarimetry is acquired. A detailed description
of this standard dual-beam polarimetry configuration can be found in
Berry \& Gledhill \shortcite{berry03}. Basically, two orthogonally
polarized images are simultaneously recorded. The subtraction of these
two images leads to the suppression of unpolarized light (mainly from
the central star), and to the highlighting of polarized light (e.g.
scattered light from the disc).

We observed a total of 11 objects. Since most of our targets are
bright (K$<8$), several short exposures ($<1$s) were co-added, in
order to overcome saturation problems. Typical integration times
ranged from 315 to 475 seconds. A non-destructive readout mode was
used (ND+STARE). Point Spread Function (PSF) and flux calibrators were
regularly interspersed, and polarized standards from the UKIRT
calibration catalogue were also observed. In order to minimise the
effects of pixel to pixel variations in the detector on derived
polarization values, each measurement was repeated at 3 different
array (jitter) positions.


\section{Data reduction}

All data were identically reduced using the STARLINK data-reduction
pipeline ORAC-DR\footnote{\tt http://www.oracdr.org/}. Dark
subtraction and flat-fielding are carried out, and the dual-beam
imaging polarimetry package POLPACK \cite{berry03} was then used to
perform component image alignment and to combine the data to form
resultant I, Q and U Stokes images. Propagation of the variance
estimates from the raw data through the calculation provides errors on
the final I, Q and U images, and on the derived polarized quantities P
and PI (degree of polarization and polarized flux, respectively).

The values of P presented in this work correspond to the degree of
polarization averaged over 0.36 arcsecond square bins (3$\times$3
pixels), after applying a 2 $\sigma$ cut. In this way, only
measurements with an error in the degree of polarization of less than
one-half of the polarization are considered.

\section{Results} 
\subsection{J- band imaging polarimetry and photometry}  \label{resultsj}

J- band photometry is presented in Table~\ref{jresults}, together with
2MASS photometry. The photometric values derived from our measurements
are generally in good agreement with the 2MASS values. HD~143006's J-
band magnitude shows a 0.23 magnitude deviation with respect to
previous J- band photometry, suggesting that the object has brightened
since the 2MASS measurements. HD~169142's J- band magnitude is also
fainter than the 2MASS photometry, but is consistent with J- band
measurements from Sylvester et al. \shortcite{syl96}.\\

Table~\ref{jpolresults} presents measured integrated polarizations for
our targeted stars. In the case of resolved centrosymmetric patterns
(as in the cases of TW~Hya and HD~169142), the integrated
polarization will appear less than the actual resolved polarizations,
since integrating the polarization vectors over all orientations can
average to zero. As an example, TW Hya's maximum resolved polarization
is $15~\pm~0.5~\%$ at $1.6$~arcseconds from the star, while its
integrated polarization is only $0.233~\pm~0.025~\%$. The J- band
instrumental polarization was estimated from observations of
unpolarized standards, which showed percentage polarization levels of
less than $0.5$ percent. These values represent the real precision of
our polarimetric measurements (mostly residuals from the alignment
procedure), and should again not be confused with the statistical
errors on the integrated polarizations quoted in
Table~\ref{jpolresults}.\\


Targets for which we detect evidence of polarizing circumstellar
material are discussed individually below (TW Hya, HD~169142,
HD~150193 and HD~142666). The remaining stars from our sample do not
show any evidence of dusty polarizing discs. This suggests that their
polzarization levels are too faint to be detected by our observations,
as exemplified by the non-detection of the disc around HD~141569. The
intensity distribution of HD~141569's ring-like disc is known to peak
at $3.3$~arcseconds from the star \cite{aug99}, a scale resolvable
with UKIRT. However our J-band polarimetric observations were not able
to detect it. Combining the HST J-band coronographic observations of
HD~141569's disc \cite{aug99} and our polarimetric data, we derive
upper-limits on the polarization level of HD~141569's disc. From this
we conclude the disc is so faint that even if it was $100\%$ polarized
it would still fall below our detection limit. Augereau et al.'s
\shortcite{aug99} HST observations showed that the surface brightness
of HD~141569's disc peaks at $0.199\,\pm\,1.2$ mJy arcsec$^{-2}$. This
is one order of magnitude fainter than that of TW Hya's disc
($5.7\,\pm\,1.4$ mJy arcsec$^{-2}$; Weinberger et
al. \shortcite{wein02}), and could explain why the disc around TW Hya
could be seen in polarized light, while the one around HD~141569 was
not.\\


J-band imaging polarimetry of four of our eleven targets is presented
in Figures ~\ref{fig0},~\ref{fig2},~\ref{fig4} and~\ref{fig5} using
the same format for each target. Polarization vector maps are
superimposed upon total intensity contours. Total intensity
grey-scales are also plotted, with the purpose of highlighting the
position of the central source. The polarization vectors are oriented
parallel to the E vector, with their length proportional to the degree
of linear polarization, as indicated by the scale vector on each
diagram.\\

\begin{description}

\item {\bf{TW Hya}}:\\


Our observations show a centro-symmetric polarization pattern (Figure
\ref{fig0}, left panel), revealing the signature of a scattering disc
seen near to face-on. The polarization values increase with increasing
radius, up to $\sim15~\%$ at $1.6$ arcseconds, although these have
larger uncertainties than those close in. We note the $\sim15~\%$
percentage polarization values we measure at $1.6$ arcseconds include
direct unpolarized flux from the star in the estimate of the disc's
total scattered light intensity, and so represent a lower limit on the
disc's intrinsic polarization. Please refer to
Section~\ref{discussion} and Figure \ref{fig9}, left panel, for an
estimate on the intrinsic polarization of TW Hya's disc. We estimate
the extension of the polarizing disc by comparing the polarized
intensity (PI) radial profile of TW Hya to the total intensity radial
profile of the PSF reference star HIP54690 (Figure \ref{fig0}, right
panel). Extended polarized structure can be traced from
$\sim0.4$~arcseconds up to at least $\sim1.5$~arcseconds. This is
comparable to the extension of TW Hya's polarizing disc as seen in the
Ks- band with the VLT ($1.4$~arcseconds in radius, Apai et
al. \shortcite{apai04}).\\

\item {\bf{HD~169142}}: \\


  Our UKIRT observations just marginally resolve a centro-symmetric
  pattern around HD~169142 (Figure \ref{fig2}, left panel). The
  PI radial profile of HD~169142 (Figure \ref{fig2},
  right panel) shows the detection of an extended polarizing disc from
  $\sim0.4$ to about $1.2$ arcseconds radius. Inspection of the Q-
  polarization image of HD~169142 (Figure \ref{fig3}, left panel)
  reveals a clear modulation with angle around the star (Figure
  \ref{fig3}, right panel). The modulation reveals the presence of an
  extended distribution of dust, as has already been shown by Kuhn et
  al. \shortcite{kuhn01}.\\


\item {\bf{HD~150193}}: \\

  Recent adaptive optics H-band coronographic observations of the
  binary system HD~150193 A-B ($1.1$~arcseconds separation) at the
  8.2-m Subaru telescope \cite{fuka03} have revealed the presence of a
  scattering disc around the primary star, extending from the edge of
  the coronographic mask ($0.5$~arcseconds radius, or $50$~AU) to
  about 1.3 arcseconds. An asymmetry of the disc in the direction of
  the binary companion was detected, suggesting that the companion has
  a distorting effect on the disc structure. HD~150193~A is known to
  be a A2 IV Herbig Ae star \cite{mora01}, whilst HD~150193~B has
  recently been classified as a classical T Tauri star
  \cite{bouvier01}.\\

  Our J-band observations show polarization vectors aligned at a
  position angle (PA) of approximately 57 degrees, measured from North
  to East (Figure \ref{fig4}, left panel). This sort of pattern is
  usually associated with scattering in an unresolved geometry. In
  fact, the J-band PI radial profile of HD~150193~A
  shows no extension when compared to the total intensity radial
  profile of the PSF reference star (Figure \ref{fig4}, right
  panel).\\

  HD~150193 presents the highest integrated percentage polarization of
  the sample ($3.15\%$ at PA=$57$ degrees), of which a significant
  part is thought to be interstellar in nature (HD~150193 is located
  near the edge of the $\rho$ Ophiuchus cloud). Yudin
  \shortcite{yudin00} estimated the interstellar polarization to be
  $2.5~\%$ at $21$ degrees PA in the V- band, and concluded that the
  intrinsic optical polarization of HD~150193 is $4.2\%$, at $65$
  degrees PA. By comparing our measured PA and Yudin's intrinsic PA,
  we conclude that most of the polarization we detect in the J band
  must be intrinsic to the disc, with a large amount of polarizing
  material in the line of sight of the star. We note that the
  integrated polarization that we measure around the system of $3.15\%$
  is in agreement with Whittet et al. \shortcite{whit92}, given
  systematic errors of $0.1\,-\,0.2\%$.\\

\item {\bf{HD~142666}}: \\

  This A8Ve star \cite{dun97b} was catalogued as a probable Herbig Ae
  star by Gregorio-Hetem et al. \shortcite{greg92}, yet no CS disc has
  been directly observed around it. HD~142666 shows evidence of CS
  activity; a double-peaked H$\alpha$ profile \cite{dun97b},
  photometric variability, and a large IR excess attributed to CS dust
  \cite{syl96}. Meeus, Waelkens \& Malfait \shortcite{meeus98}
  interpreted the variable photometric behaviour of HD~142666 as
  evidence of clumps of CS material crossing the line of sight of the
  star. If the CS structure is flattened, then the CS disc would be
  more likely to be orientated edge-on. Yudin, Clarke and Smith
  \shortcite{yudin99} measured 0.5\% intrinsic optical polarization,
  considerably larger than those of the Vega-like 'prototypes' (0.02\%
  and 0.007\% for $\beta$ Pictoris and Vega respectively; Leroy
  \shortcite{leroy93}).\\.

  Figure \ref{fig5}, left panel, shows our J-band imaging polarimetry
  of HD~142666. HD~142666 presents a polarization pattern with vectors
  aligned at $\sim75$ degrees PA and a $1.2~\%$ mean polarization. As
  with HD~150193, the PI radial profile of HD~142666 shows no evidence
  of extension (Figure \ref{fig5}, right panel). Combining catalogued
  polarization values and Hipparcos distances, Yudin et
  al. \shortcite{yudin99} derived a distance-polarization law that
  predicts a $0.3~\%$ contribution at V due to interstellar
  polarization, with a position angle of about $100$ degrees at
  optical wavelengths. They estimated a V-band intrinsic polarization
  of $p\,\approx 0.5~\%$ PA of $70$ degrees. The polarization we
  measure in the J- band is too high to be completely interstellar in
  nature. In addition, we measure a polarization PA similar to that of
  the intrinsic optical polarization estimated by Yudin et
  al. \shortcite{yudin99}. Yudin et al. \shortcite{yudin99} also noted
  that the polarization of HD~142666 probably increases toward IR
  wavelengths, as is confirmed by our J- band measurements. This
  implies no significant interstellar contribution to the $1.2\%$
  J-band polarization that we measure, so we conclude that the
  polarization detected is intrinsic to the system.




\subsection{H- and K- band imaging polarimetry} \label{hk}

H- and K-band imaging polarimetry was obtained for TW Hya, HD~150193
and HD~141569 on April 30th 2003. H and K photometric values are
shown in Table~\ref{jresults}, while H- and K-band polarimetric
measurements are presented in Table~\ref{jpolresults}. We note that
HD~150193 appears fainter by $\sim0.4$~magnitudes in both the H- and
K- band with respect to the 2MASS measurements.\\

TW Hya's H- and K- band imaging polarimetry are shown in
Figure~\ref{fig6}, left and right panels respectively, whilst
Figure~\ref{fig7} shows the H- and K- band imaging polarimetry of
HD~150193 (left and right panels, respectively). Alignment and
subtraction residuals dominate the central parts of the images
(radius$<0.4$~arcseconds).  Typical values for instrumental
polarization were $<0.4$ percent in both H- and K- band (as measured
from unpolarized standards).  As in the J- band, HD~141569 did not
show any evidence of extended polarized structure and is therefore not
shown. The near-IR spectral dependence of the polarization for our
three stars targeted at JHK is shown in Figure~\ref{fig8}.\\

Figure~\ref{fig9}, left panel, shows the JHK PI radial profiles of TW
Hya. Our data show that the radial distribution of the JHK PI all
follow a similar behaviour between $0.5$ and $1.3$ arcseconds from the
star, with the polarizing disc being significantly brighter at H than
it is at J and K. There is also evidence of a decrease in slope in the
PI profiles from $1.3$ to $2$ arcseconds, as has been previously seen
in direct light by Krist et al. \shortcite{krist00} and Weinberger at
al. \shortcite{wein02}. This reduction in slope at $r>1.3$ arcseconds
is clear in both H and K but is less evident at J, where a bump is
seen at r$\sim1.5$~arcseconds at very low level (see
Figure~\ref{fig0}). This is consistent with Weinberger et al.'s
\shortcite{wein02} total intensity radial profiles, in which the
change in slope is more evident at H than it is at J.

\

\section{Discussion} \label{discussion}

Our polarimetric imaging survey has successfully detected the presence
of circumstellar polarized emission around four of our eleven targets.
According to proposed CS evolutionary scenarios, both the IR excess
and polarization should decrease with time
\cite{malfait98,waelkens94,yudin99}; as a consequence of processes
such as planet formation, the discs should be cleared out and both the
near-IR excess and the polarizing effects of CS dust should
disappear. Thus, the detection of considerable amounts of polarizing
material around TW Hya, HD~150193, HD~142666 and HD~169142 is
consistent with a youthful nature. On the other hand, the
non-detection of polarized emission toward stars presenting low
IR-excesses (L$_{\rm IR}$/L$_{\rm star} < 0.1$) suggests that these
stars are at a more advanced evolutionary state, probably near to the
begining of their MS phase. In this context we would have expected to
detect polarized emission around HD~144432 and HD~143006 (which
present L$_{\rm IR}$/L$_{\rm star}$ values and near-IR excesses even
higher than those of HD~142666. The most likely reasons for not
detecting their discs polarimetrically could be that either we did not
integrate deep enough to detect the scattered light against the
stellar flux or that the disc is symmetrically distributed but
unresolved, or a combination of both effects. Dominik et
al. \shortcite{dom03} showed that the near-IR excesses of Class II
sources (such as HD~144432, HD~142666 and HD~150193) can be well
modelled if there is a large presence of micron-sized grains in the
inner 10~AU of their CS discs. It seems reasonable to deduce that the
near-IR scattering discs around HD~144432, HD~143006 and HD~142666 are
too small to be resolved by UKIRT. In fact, recent VLT mid-IR
interferometric observations have successfully measured the radius of
HD~144432's disc to be of $0.014$ arcseconds \cite{leinert04}, far too
small to be resolved by our UKIRT observations.\\

Apai et al. \shortcite{apai04} proposed that TW Hya's disc Ks-band
polarized intensity profile and the J- and H-band total intensity
radial profiles have a similar behaviour, concluding that the
percentage of polarization is nearly constant between $0.9$ and $1.4$
arcseconds from the star. As reported in Section~\ref{hk}, our
observations show that in this region the JHK PI radial profiles are
indeed very similar. By combining our PI radial profiles and the total
intensity radial profiles from HST \cite{wein02}, we derive TW Hya's
disc J- and H-band intrinsic percentage polarization as a function of
radius (Figure~\ref{fig9}, right panel). We find P$_{\rm J}$ and
P$_{\rm H}$ are roughly constant between $0.9$ and $1.3$ arcseconds,
with P$_{\rm H}$ being higher than P$_{\rm J}$ at all radii. In this
region, the percentage polarization has peaks at $27.6 \pm 6.1$ at J
and $31.3 \pm 1.5$ at H, both at a radius of $1.2$ arcseconds. Beyond
$1.4$~arcseconds P$_{\rm J}$ falls-off, while P$_{\rm H}$ rises with
increasing offset from the star. This is a consequence of the steeper
J-band PI profile, as the Weinberger et al. \shortcite{wein02} total
intensity profiles for the $1.1$ and $1.6$~$\mu$m NICMOS filters have
the same $r^{-2.6\,\pm\,0.1}$ fitted slope. However, the NICMOS
$1.6$~$\mu$m surface brightness deviates from this single power-law
beyond $1.4$~arcseconds, suggesting that our P$_{\rm H}$ may be
overestimated beyond this radius. Nevertheless, we can conclude that
our observations, in conjunction with direct light images from the HST
\cite{wein02}, indicate that the disc scattering and polarizing
efficiencies peak in the H-band, suggesting there is a substantial
population of sub-micron sized particles present in the disc
\cite{bohren83}. This would argue against larger grains of the type
suggested by Roberge, Weinberger \& Malumuth \shortcite{roberge05},
although this needs to be investigated more thoroughly by future
modelling.\\





Our imaging polarimetry for HD~169142 confirms the presence of an
extended CS structure. Both the Q- and U- J-band images showed
modulation with angle as was already noted by Kuhn et
al. \shortcite{kuhn01}, suggesting that the disc is oriented near to
pole-on. The disc-size of $1.2$~arcseconds we measure ($174$~AU at
HD~169142's distance) is $0.3$~arcseconds smaller than the previous
polarized-disc extension measured by Kuhn et
al. \shortcite{kuhn01}. We note that the effective exposure time used
by Kuhn et al. \shortcite{kuhn01} was $2.2$ times larger than the one
we used, which may explain the difference in HD~169142's measured disc
extension between these two UKIRT observations. Kuhn et
al. \shortcite{kuhn01} also detected a scattering inhomogeneity within
the CS cloud, but this was not evident in our data. It is interesting
to note that HD~169142 has been proposed as a Group I source
(following the Meeus et al. \shortcite{meeus01} classification) and
that its SED has been successfully modelled with a flaring-disc
geometry \cite{dom03} and outer disc-radius of $100$~AU.  \\

Based on the H-band disc-radius of $1.3$~arcseconds measured
coronographically with Subaru by Fukagawa et al. \shortcite{fuka03},
we would have expected to resolve the disc around HD~150193, at least
in our H- and K-band observations. However we find no signs of
extension when comparing our PI radial profiles of HD~150193 with the
total intensity radial profiles of the PSF calibrator star. We believe
our non-detection of the extended disc is due to sensitivity effects,
since the Subaru observations carried out with a nine times longer
integration time and a four times larger telescope collecting
area. Even though we failed to detect the disc reported by Fukagawa et
al. \shortcite{fuka03}, we believe there must still be an axisymmetric
structure causing the relatively large linear polarization toward this
star. The alignment of the polarization vectors at $57$ degrees PA is
likely to be intrinsic, and not associated with interstellar
polarization (as discussed in Section~\ref{resultsj}). To produce the
aligned vectors such structure must be (i) optically thick at J, (ii)
inclined to the plane of the sky (i.e. not face on) and (iii) too
small to resolve with UKIRT. So this could be an inner, optically
thick component of HD~150193's disc. We also note that HD~150193 shows
the steepest spectral dependence of polarization among our three JHK
targeted stars, with its polarization PA being constant within the
errors at all wavelengths (as can be seen in both Figure~\ref{fig7}
and in the values presented in Tables~\ref{jpolresults}). Adopting the
binary PA of $225$~degrees measured by Fukagawa et
al. \shortcite{fuka03}, we note that there is an apparent alignment
between polarization PA (57 degrees) and the axis between the two
stars, although this could be just a coincidence. Nevertheless, this
behaviour has already been reported in the V-band by Maheswar, Manoj
\& Bhatt \shortcite{mahe02}, who suggested that it is common to
several other HAeBe binary systems.\\

We have found for HD~142666 similar evidence of a CS scattering
structure to that for HD~150193. Along with HD~150193, HD~142666
exhibits a large value of linear polarization compared to the rest of
our sample and also presents aligned vectors suggesting that the CS
disc was unresolved. The similarity between both stars in terms of
their infrared excesses and their polarization makes it plausible to
suggest a similar structure in their discs. This is consistent with
the SED modelling carried out by Dominik et al. \shortcite{dom03}, in
which both discs are estimated to be comparable in size and mass.\\

Despite the limited size of our sample, we note that our polarimetric
observations detected the presence of extended polarizing discs only
around systems that have had their SEDs fitted with a flared disc
geometry \cite{calvet02,dom03}. This points toward an observational
selection effect predicted by Whitney \& Hartmann
\shortcite{whitney92}, in which Monte Carlo scattered light models
show that a flaring disc scatters up to two orders of magnitudes more
light than a flat disc of similar size and mass. Detailed
scattered-light modelling of our polarimetric data is, however,
required to constrain the physical parameters of the detected CS
discs.\\

\section*{Acknowledgements}

ASH carried out this work whilst being funded as part of the PPARC
Gemini - Fundaci\'{o}n Andes UK/Chile studentship programme. We thank
Dr Malcolm Currie for very useful help with aspects of the data
reduction. This work made use of the Starlink computing network,
funded by the UK Particle Physics and Astronomy Research Council
(PPARC). The UKIRT is operated by the Joint Astronomy Centre on behalf
of PPARC, the Netherlands Organization for Pure Research, and the
National Research Council of Canada. This work made use of the SIMBAD
database and other facilities operated at CDS, Strasbourg, France, and
of the 2MASS point-source catalogue available at the NASA/IPAC
Infrared Science Archive, which is operated by the Jet Propulsion
Laboratory, California Institute of Technology, under contract with
the National Aeronautics and Space Administration. We especially thank 
the anonymous referee for helpful suggestions that greatly improved the 
article.


\begin{center}
\begin{table*}
\caption{Programme stars observed}
\begin{tabular}{@{}cccccccc}
Name     &Sp class& L$_{\rm IR}$/L$_{\rm \star}$&Distance [pc]&V\\
\hline

HD~ 98800 &K5Ve&8.4$\cdot10^{-2}$&17&8.86\\
HD~123160 &G5V&4.4$\cdot10^{-3}$&15.7&8.62\\
HD~123356 &G1V&5$\cdot10^{-2}$&41&9.7 \\
HD~141569 &A0Ve&8.4$\cdot10^{-3}$&99&7.0 \\  
HD~142666 &A8Ve&0.34&114&8.65 \\
HD~143006 &G5Ve &0.56&82&10.21\\
HD~144432 &A9/F0Ve&0.48&200&8.17 \\
HD~145263 &F0V&2$\cdot10^{-2}$&116&8.95 \\
HD~150193 &A1V&0.37&150&8.88 \\ 
HD~169142 &A5Ve&8.8$\cdot10^{-2}$&145&8.15 \\
TW  Hya  &K7 Ve &0.25  & 56&11.1\\ 
\end{tabular}
\label{sample} 
\medskip

 Spectral classifications and IR excesses are from Sylvester et al.
 \shortcite{syl96}, Dunkin, Barlow \& Ryan \shortcite{dun97a} and
 Sylvester \& Mannings \shortcite{syl00}. An `e' after the luminosity
 class indicates the presence of H$\alpha$ emission.  HD~141569,
 HD~123356, HD~144432, HD~150193 and HD~98800 have binary companions
 at separations of 6.8, 2.2, 1.2, 1.1 and 0.8~arcseconds respectively
 \cite{aug99,syl00,perez04,mora01,prato01}. Distances are from the
 Hipparcos catalogue \cite{perry97}.
\end{table*}
\end{center}

\begin{center}
\begin{table*}
\caption{Photometric measurements}
\begin{tabular}{@{}cccccc}
name&Filter&Magnitude&Magnitude& FWHM  &FWHM\\
    &      &         &2MASS    &       &reference\\       
    &      &         &         & [arcseconds]&[arcseconds]\\
\hline

HD~ 98800 &J&6.36 $\pm$ $0.02$&6.39 $\pm$ $0.02$&0.84 & 0.87     \\
HD~123160 &J&5.78 $\pm$ $0.03$&5.81 $\pm$ $0.02$& 1.30 &1.23      \\
HD~123356 &J&5.94 $\pm$ $0.02$&5.89 $\pm$ $0.02$&0.94 &0.93    \\
HD~141569 &J&6.88 $\pm$ $0.02$&6.87 $\pm$ $0.03$&0.85&1.09         \\
HD~141569 &H&7.05 $\pm$ $0.02$&6.86 $\pm$ $0.02$&0.41 &0.45   \\
HD~141569 &K&6.84 $\pm$ $0.02$&6.82 $\pm$ $0.02$&0.38 &0.39   \\
HD~142666 &J&7.32 $\pm$ $0.02$&7.35 $\pm$ $0.02$&0.93&0.92        \\
HD~143006 &J&8.12 $\pm$ $0.03$&8.35 $\pm$ $0.02$&0.86&0.95        \\
HD~144432 &J&7.13 $\pm$ $0.03$&7.09 $\pm$ $0.02$&0.85&0.84       \\
HD~145263 &J&8.10 $\pm$ $0.02$&8.08 $\pm$ $0.02$&0.69&0.67         \\
HD~150193 &J&6.94 $\pm$ $0.02$&6.94 $\pm$ $0.02$&0.66&0.65         \\
HD~150193 &H&6.63 $\pm$ $0.02$&6.21 $\pm$ $0.02$&0.53 & 0.51  \\
HD~150193 &K&5.92 $\pm$ $0.02$&5.47 $\pm$ $0.02$&0.39 & 0.37  \\
HD~169142 &J&7.44 $\pm$ $0.02$&7.31 $\pm$ $0.02$&0.66&0.58         \\
TW Hya    &J&8.16 $\pm$ $0.02$&8.21 $\pm$ $0.02$&0.82&0.88          \\
TW Hya    &H&7.58 $\pm$ $0.02$&7.55 $\pm$ $0.02$&0.51 &0.46   \\
TW Hya    &K&7.29 $\pm$ $0.02$&7.29 $\pm$ $0.02$&0.53 &0.49   \\

\end{tabular}
\label{jresults}
 
\medskip
Photometry was performed using 3.5 arcsecond apertures and calibrated
using observed UKIRT standards. Photometric magnitudes from the 2MASS
point-source catalogue have been included. FWHMs measured both in the
targeted stars and repective PSF references are presented.
\end{table*}
\end{center}

\begin{center}
\begin{table*}
\caption{Polarimetric measurements}
\begin{tabular}{@{}cccc}
name&Filter&P [\%]&$\theta$\\
\hline

HD~ 98800&J& 0.21 $\pm$ $0.02$ & -70 $\pm$ $11$  \\
HD~123160&J& 0.12 $\pm$ $0.01$ &  80 $\pm$ $13$     \\
HD~123356&J& 0.26 $\pm$ $0.01$ &  84 $\pm$ $9$  \\
HD~141569&J& 0.55 $\pm$ $0.02$ &-77  $\pm$ $9$    \\
HD~141569&H& 0.22 $\pm$ $0.02$ &-71  $\pm$ $8$    \\
HD~141569&K& 0.08 $\pm$ $0.02$ &-59  $\pm$ $10$    \\
HD~142666&J& 1.32 $\pm$ $0.02 $ & 75 $\pm$ $3$    \\
HD~143006&J& 0.17 $\pm$ $0.02$ &  6  $\pm$ $12$   \\
HD~144432&J& 0.49 $\pm$ $0.02 $ & 3  $\pm$ $8$   \\
HD~145263&J& 0.37 $\pm$ $0.03$ & 23  $\pm$ $11$    \\
HD~150193&J& 3.14 $\pm$ $0.02$ & 57  $\pm$ $4$    \\
HD~150193&H& 2.04 $\pm$ $0.01$ & 52  $\pm$ $5$    \\
HD~150193&K& 1.41 $\pm$ $0.03$ & 48  $\pm$ $6$    \\
HD~169142&J& 0.22 $\pm$ $0.03$ &-37  $\pm$ $6$    \\
TW Hya   &J& 0.23 $\pm$ $0.02$ & 51  $\pm$ $6$     \\
TW Hya   &H& 0.18 $\pm$ $0.02$ & 56  $\pm$ $7$     \\
TW Hya   &K& 0.10 $\pm$ $0.02$ & 50  $\pm$ $8$     \\

\end{tabular}
\label{jpolresults}
 
\medskip

Integrated percentage polarizations were computed using the
polarimetry tool-box in GAIA$^{\dag}$, which allows vectors within a
chosen aperture to be selected for integration {\dag}({\tt
http://www.starlink.rl.ac.uk/}). This calculates the mean Stokes Q and
U values and returns the polarization integrated over the annular
region defined by $0.4<$radius$<3.5$~arcseconds. This excludes the
inner regions which are usually dominated by alignment and subtraction
residuals. Polarization position angles are in degrees and measured
from North to East.

\end{table*}
\end{center}


\begin{figure*}
\centerline{\psfig{file=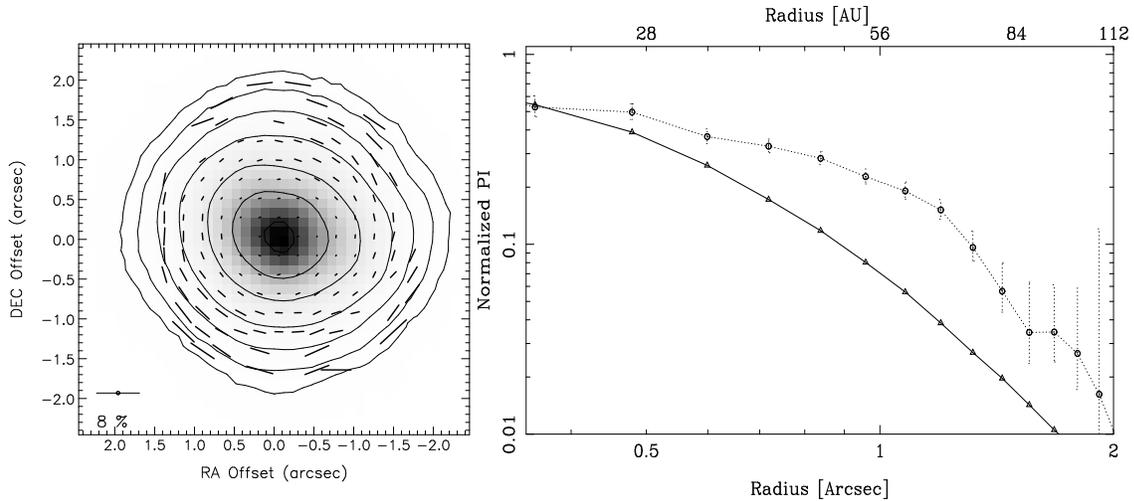}}
\caption{\small {\it{ Left panel:}} J-Band imaging polarimetry of TW
  Hya. Polarization vector maps are superimposed upon total intensity
  maps, showing both contours and grey-scale images. Polarization
  measurements have been binned in $3\times3$ square bins and a
  $2\sigma$ cut applied to avoid spurious values. North is up and East
  is to the left. {\it{ Right panel:}} Normalized radial polarized
  intensity distribution around TW Hya (dotted line) compared to the
  total intensity radial profile of the PSF calibrator star HIP~54690
  (solid line). The data have been binned over 1-pixel width rings and
  the error bars correspond to the statistical errors of each
  bin. Both x- and y- axis are plotted on logarithmic scale.}
\label{fig0}
\end{figure*}


\begin{figure*}
\centerline{\psfig{file=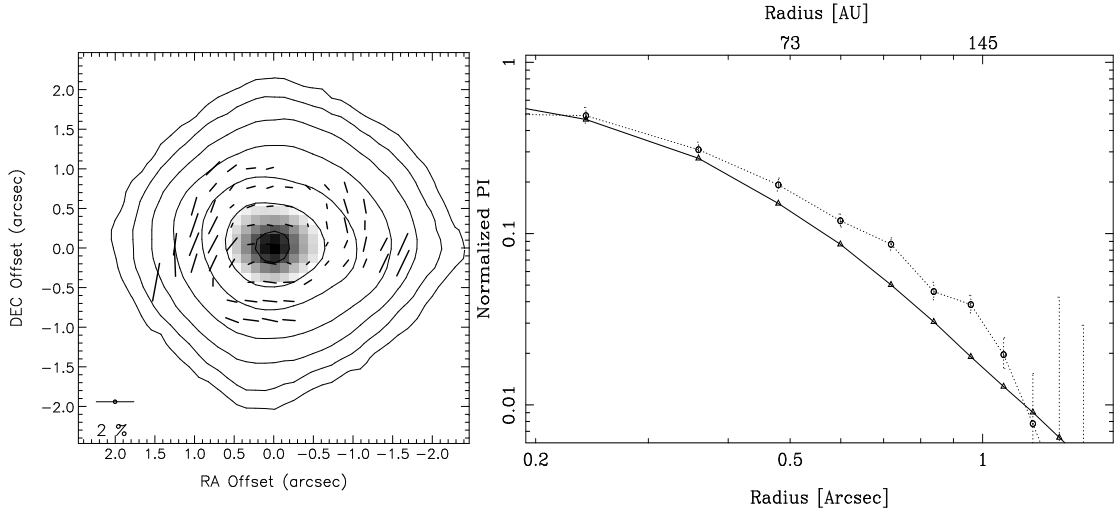}}
\caption{\small{\it{ Left panel:}} J-band imaging polarimetry of
HD~169142, presented in the same format as for Figure 1. The
polarization vector map of HD~169142 suggests the marginal detection
of a centro-symmetric pattern. {\it{ Right panel:}} The J-band
PI radial profile for HD~169142 (dotted line). The
solid line corresponds to the total intensity radial profile of the
PSF reference star, FS~140. Azimuthal values have been binned over
1-pixel width concentric rings. The error bars represent the
statistical errors. Both x- and y- axis are plotted on logarithmic
scale.}
\label{fig2}
\end{figure*}

\begin{figure*}
\centerline{\psfig{file=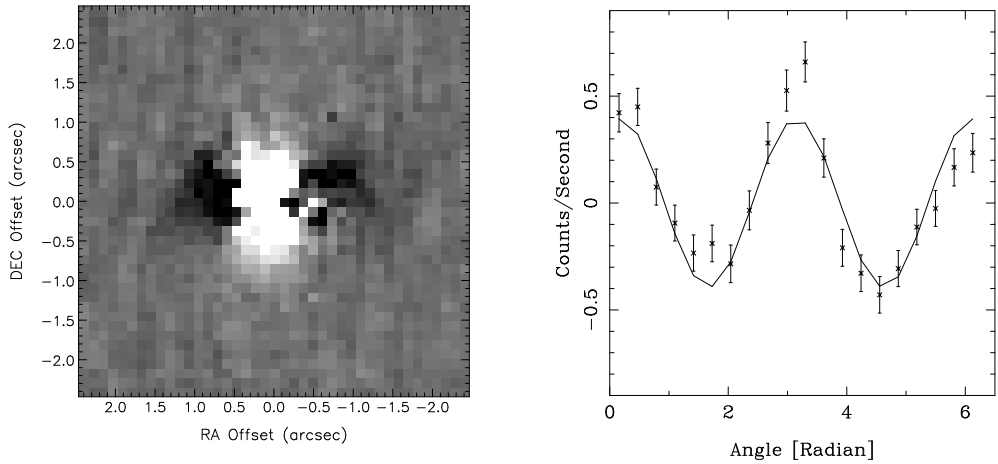}}
\caption{ Q- Stokes image of HD~169142 (left panel).
HD~169142's Q- Stokes image shows sinusoidal modulation with
angle (right panel). The crosses mark the data points binned over 18
degree bins, centred on HD~169142, using an annulus extending between
0.6 and 2.2 arcseconds. The solid curve represents the best sinusoidal
fit to the data.  }
\label{fig3}
\end{figure*}
\end{description}

\begin{figure*}
\centerline{\psfig{file=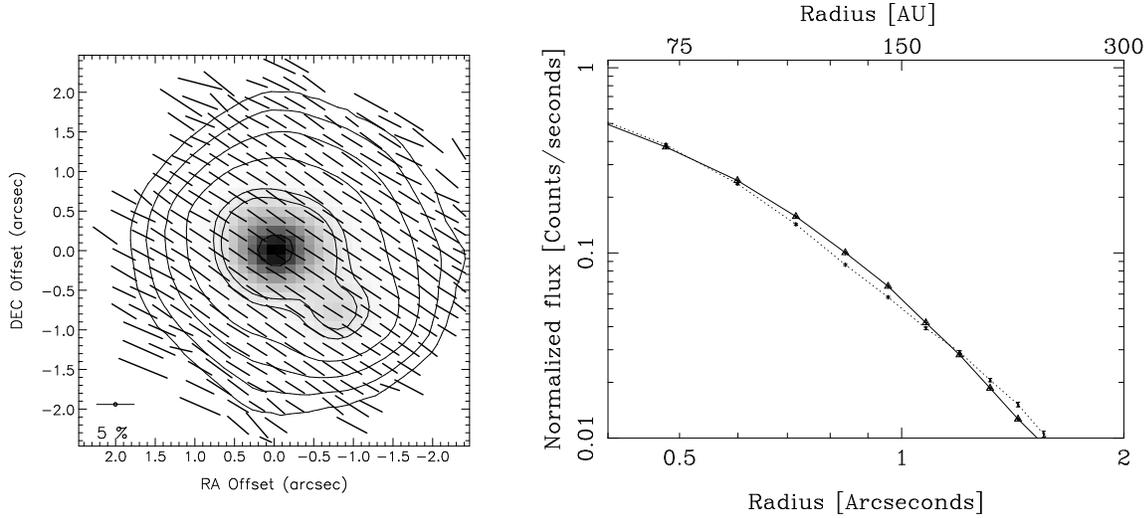}}
\caption{\small{\it{ Left panel:}} J-Band imaging polarimetry of
HD~150193. Polarization vectors are superimposed upon total intensity
maps shown as both grey-scale and contour images. The polarization
measurements have been binned in $3\times3$ square bins and a
$2\sigma$ cut applied to avoid spurious values. North is up and East
is to the left. {\it{ Right panel:}} HD~150193~A's PI
radial profile (dotted line) versus the total intensity radial profile
of the PSF reference star HIP~80425 (solid line). The region
containing HD~150193~B was excluded when computing HD~150193~A's
PI radial profile. Both x- and y- axis are plotted on
logarithmic scale.}
\label{fig4}
\end{figure*}

\begin{figure*}
\centerline{\psfig{file=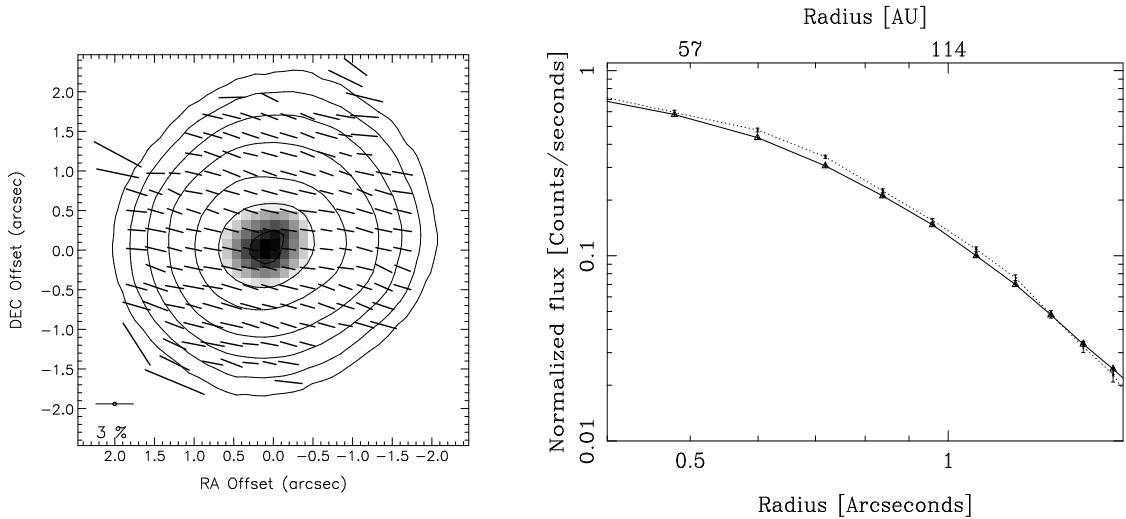}}
\caption{\small{\it{ Left panel:}} J-Band imaging polarimetry of
HD~142666. North is up and East is to the left.  {\it{Right panel:}}
HD~142666 's PI radial profile (dotted line) versus
the total intensity radial profile of the PSF reference star HIP~77815
(solid line). Both x- and y- axis are plotted on logarithmic scale.}
\label{fig5}
\end{figure*}

\begin{figure*}
\centerline{\psfig{file=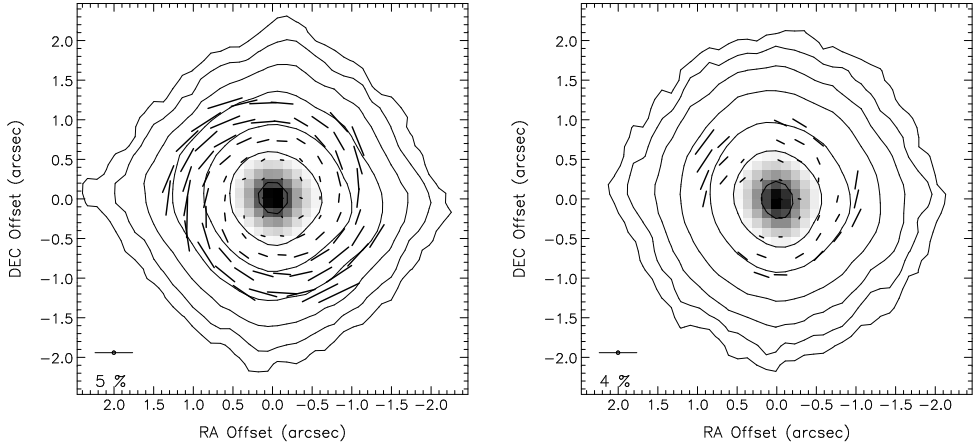}}
\caption{H-band (left) and K-band (right) imaging polarimetry of TW
 Hya. Polarization vector maps are superimposed upon total intensity
 maps, shown as both contour and grey-scale images. North is up and
 East is to the left.}
\label{fig6}
\end{figure*}

\begin{figure*}
\centerline{\psfig{file=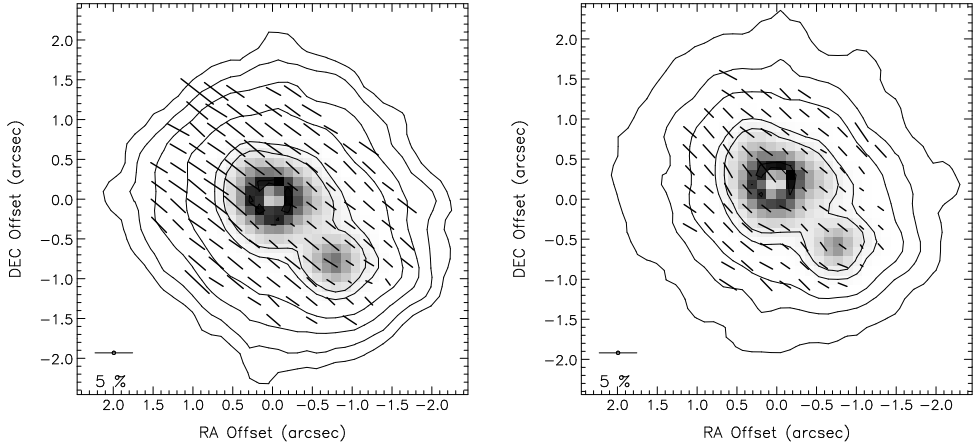}}
\caption{H-band (left) and K-band (right) imaging polarimetry of
HD~150193. Polarization vector maps are superimposed upon total
intensity contour and grey-scale images. North is up and East is to
the left.}
\label{fig7}
\end{figure*}

\begin{figure*}
\centerline{\psfig{file=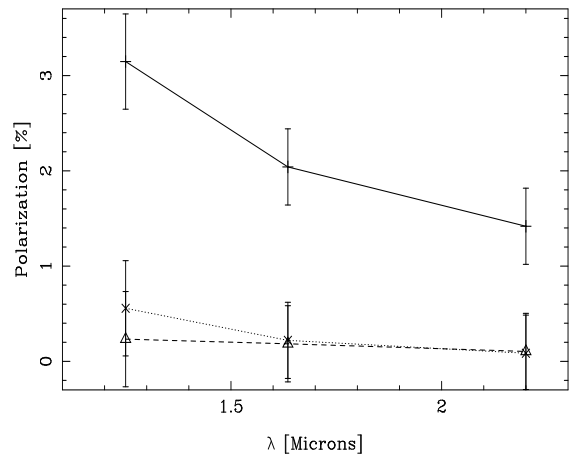}}
\caption{Near-IR polarization spectral dependence for HD~150193,
HD~141569 and TW Hya (solid, dotted and dashed lines
respectively). The error bars correspond to the instrumental errors.}
\label{fig8}
\end{figure*}

\begin{figure*}
\centerline{\psfig{file=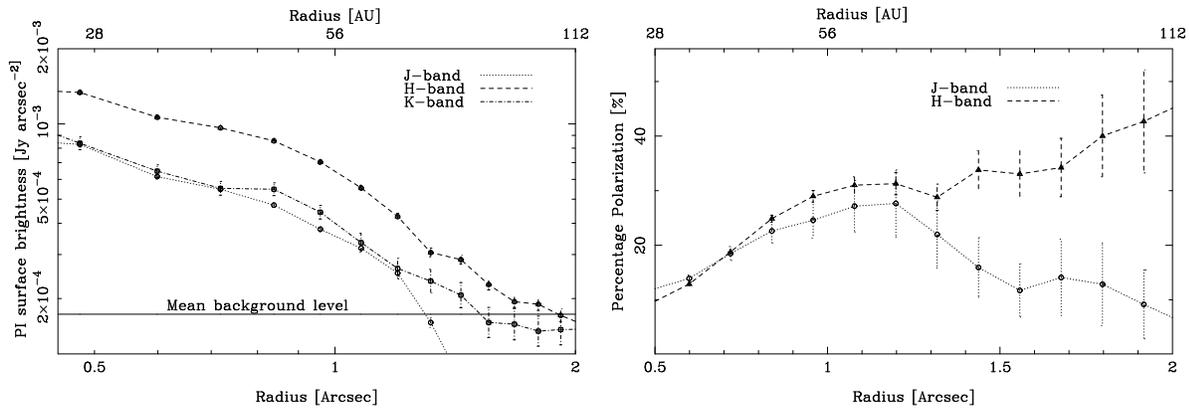}}
\caption{\small{\it{ Left panel:}} J-, H- and K-band polarized
intensity radial profiles of TW Hya (dotted, dashed and solid lines
respectively). Both x- and y- axis are plotted on logarithmic
scale. \small{\it{ Right panel:}} J- (circles) and H-band (triangles)
intrinsic percentage polarization of TW Hya's disc, derived using
total intensity coronographic imaging from HST (Weinberger et
al. \shortcite{wein02}) and our UKIRT imaging polarimetry. The error
bars correspond to derived statistical errors.}
\label{fig9}
\end{figure*}


\label{lastpage}

\end{document}